\documentclass[prl,twocolumn,notitlepage,superscriptaddress,longbibliography]{revtex4-1}
\usepackage{amsmath}
\usepackage{amssymb}
\usepackage{mathrsfs}
\usepackage[english]{babel}

\usepackage[unicode=true,colorlinks=true,citecolor=blue,urlcolor=blue]{hyperref}

\usepackage{bm}
\usepackage{color}
\usepackage{epsfig}

\usepackage[normalem]{ulem}

\renewcommand {\i}{{\rm i}}
\renewcommand {\phi}{{\varphi}}

\begin{document}
\title{
Cavity optomechanics in ultrastrong light matter coupling regime: Self-alignment and collective rotation mediated by Casimir torque
}

\author{Denis Ilin}
\affiliation{Department of Physics and Technology, ITMO University, St. Petersburg, 197101, Russia}

\author{I.~V.~Tokatly}\affiliation{Nano-Bio Spectroscopy Group and European Theoretical Spectroscopy Facility (ETSF), Departamento de Polímeros y Materiales
Avanzados: Física, Química y Tecnología, Universidad del País Vasco, Avenida Tolosa 72, E-20018 San Sebastián, Spain}
\affiliation{IKERBASQUE, Basque Foundation for Science, 48009 Bilbao, Spain}
\affiliation{Donostia International Physics Center (DIPC), E-20018 Donostia-San Sebastián, Spain}

\author{Ivan Iorsh}
\affiliation{Department of Physics and Technology, ITMO University, St. Petersburg, 197101, Russia}

\begin{abstract}
We theoretically consider an ensemble of quantum dimers placed inside an optical cavity. We predict two effects: first, an exchange of angular momentum between the dimers mediated by the emission and re-absorption of the cavity photons leads to the alignment of dimers. Furthermore, the optical angular momentum of the vacuum state of the chiral cavity is transferred to the ensemble of dimers which leads to the synchronous rotation of the dimers at certain levels of light-matter coupling strength. 
\end{abstract}
\date{\today}

\maketitle 
\section*{Introduction}

Rapidly emerging field of cavity quantum materials~\cite{schlawin2022cavity} explores the routes to tailor the low energy electronic properties of cavity embedded low-dimensional materials via engineering of the cavity electromagnetic vacuum fluctuations. The substantial developments in this field were dictated by the tremendous progress in photonic  technologies, which allowed to routinely fabricate microcavities with large quality factors and extremely small mode volumes, and therefore, deeply subwavelength field localization. This in turn enabled the regime of the ultrastrong light-matter coupling~\cite{Kockum2019}, in which the characteristic energy of light-matter interaction becomes comparable to the cavity photon energy leading to the drastic increase in the role of the vacuum fluctuations. Utrastrong coupling was predicted to induce various cavity mediated phase transitions ~\cite{thomas2019exploring,curtis2019cavity,sentef2018cavity,schlawin2019cavity,li2020manipulating,ashida2020quantum,PhysRevLett.125.257604, wang2019cavity} and to substantially modify the chemical reactions~\cite{martinez2018can, Ebbesen2016,ebbesen2016hybrid,GarciaVidal2021,ahn2023modification}.

The mechanical degrees of freedom play crucial role in the physics of ultrastrong light-matter coupling due to many reasons. First of all, coupling of light to mechanical vibrations in organic molecules can reach ultrastrong regime even in relatively low quality factor cavities~\cite{hertzog2021enhancing, thomas2021cavity}. Moreover, recently the ultrastrong coupling of microwave mechanical vibrations in nanomechanical systems and light was realized in microcavity systems~\cite{fogliano2021mapping,PhysRevLett.131.067001}. In these set-ups, strong non-linear optomechanical response emerges already at the single photon pump level. There were several theoretical predictions on the quantum-correlated mechanical motion of atoms in the cavities under weak optical pump~\cite{chang2013self, manzoni2017designing, iorsh2020waveguide, sedov2020chiral}. Finally, as it is known from the seminal work on the Casimir effect~\cite{casimir1948influence} electromagnetic vacuum fluctuations \textit{per se}  produce mechanical force. It was recently shown that this force may be utilized for the self-assembly of polarizeable nano-objects inside microcavities~\cite{munkhbat2021tunable}.

\begin{figure}[t]
\label{fig:entrX}
\centering
\includegraphics[width=0.4\textwidth]{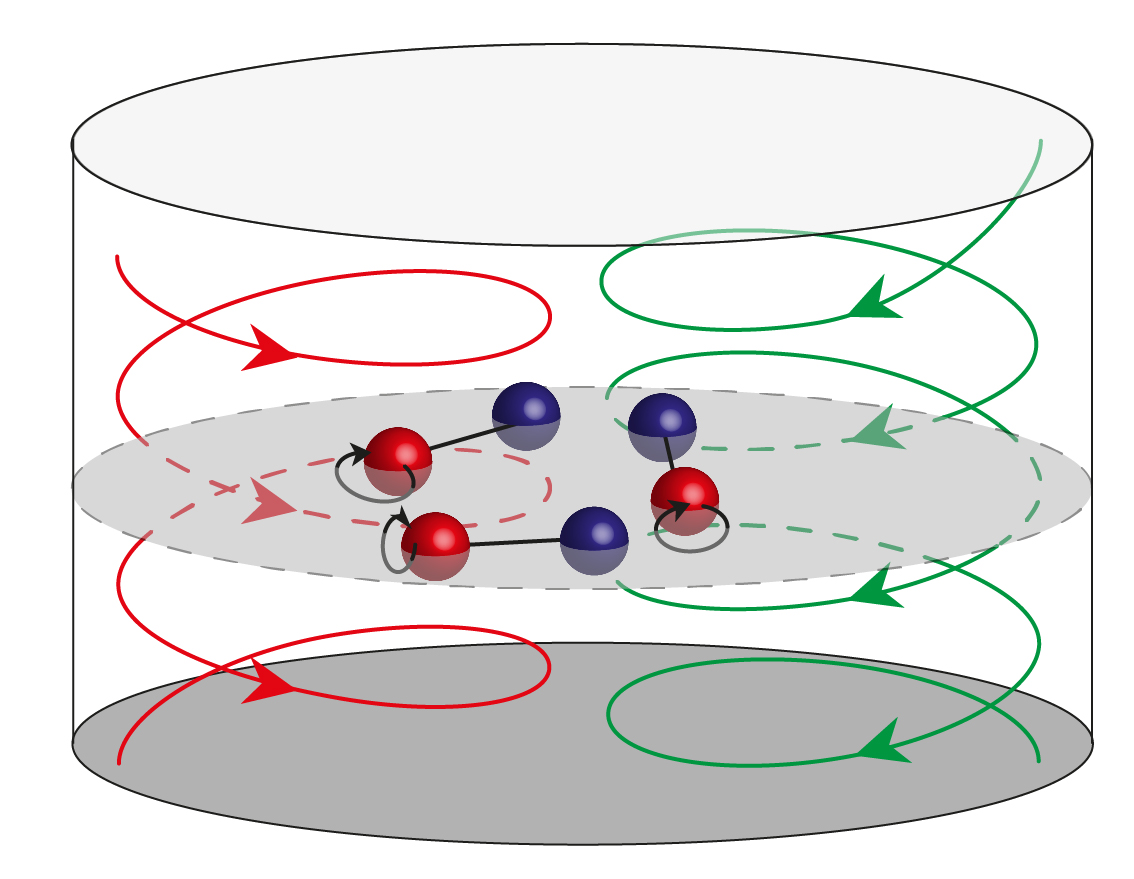}
\caption{ The geometry of the system. An ensemble of dimers is placed inside a chiral cavity. The dimers may exchange angular momentum via emission and re-absorption of cavity photons. Due to the breaking of time-reversal symmetry, cavity modes of opposite helicity have different energies, which leads to non-vanishing optical angular momentum in the ground state.
}\label{fig:entrX}
\end{figure}
Cavity induced mechanical forces originating between nano-objects placed inside the cavity can be regarded as the momentum exchange carried by the electromagnetic vacuum fluctuations. As it is known, electromagnetic field acting on a single scatterer lacking cylindrical symmetry induces both optical force and optical torque. This immediately suggests that the exchange of vacuum electromagnetic fluctuations between asymmetric objects can induce aligning or anti-aligning force. This aligning torque is called Casimir torque in analogy with Casimir force induced by vacuum electromagnetic field fluctuations, and has been recently observed experimentally~\cite{somers2018measurement}. Moreover, in the case when vacuum electromagnetic fluctuations in a cavity carry non-vanishing angular momentum, it can be transferred to the scatterer and lead to the rotation of the scatterer. Non-vanishing optical angular momentum in the ground state may appear in chiral cavities with broken time-reversal symmetry. If we consider a Fabry-Perot cavity with the mirrors made of ferromagnetic material, the optical  modes with opposite circular polarizations will have different frequencies with the energy splitting proportional to magnetization.  Ultrasong light matter coupling in chiral optical cavities~\cite{hubener2021engineering} is currently a rapidly developing area of research where multiple novel effects have been proposed ranging from cavity induced anomalous Hall effect~\cite{PhysRevB.106.205114} to the emergence of peculiar multiphoton correlated states~\cite{kurilovich2022stabilizing}.

In this Paper, we consider an array of quantum dimers placed inside a chiral optical cavity as shown in Figure~\ref{fig:entrX}. Due to the breaking of time-reversal symmetry, the vacuum state of electromagnetic field carries non-vanishing angular momentum which can be partially transferred to the ensemble of dimers. Moreover, due to the lack of cylindrical symmetry in the dimers, they can exchange angular momentum via emission and re-absorption of cavity photons. In  what follows, we provide a quantitative theoretical description of these two effects and show that it leads to the cavity mediated alignment of dimers and may induces synchronous rotation of the dimers in the cavity.

\section{Model}
We consider an ensemble of $N$ dimers in a cavity. Each dimer is characterized by the resonant frrequency $\Delta$, and light-matter coupling strength $g$. We assume that all dimers are located at a single plane parallel to the cavity mirrors. Orientation of the each dimer is defined by the angle $\varphi_i$. We also allow the dimers to rotate in the plane of the cavity mirrors with moment of inertia $J$. The full Hamiltonian of the system written in the dipole gauge can be written as
\begin{equation}
    \begin{aligned}\label{eq:trunc}
        &\hat{H}_{dip}=\frac{\Delta}{2}\sum\limits_{i=1}^{N}\hat{\sigma}_{z_i} - \sum\limits_{i=1}^{N}\frac{\partial^2_{\varphi_i}}{2J} +\frac{\omega^2}{2}\mathbf{q}^2 \\ &+ \frac{1}{2}\left({\mbox{\boldmath$\pi$}} +B[\textbf{e}_z\times \mathbf{q}]- \frac{g}{2\sqrt{N}}\sum\limits_{i=1}^{N}\hat{\sigma}_{x_i}\mathbf{n}\right)^2, 
    \end{aligned}
\end{equation}
where $\mathbf{q}=(q_x,q_y)$ is the canonical coordinate of the electromagnetic field in the cavity, $\mathbf{\pi}$ is the conjugate momentum. The second term in Eq.~\eqref{eq:trunc} corresponds to the energy of the mechanical motion of the dimers. $\mathbf{n}=(\cos\varphi_i,\sin\varphi_i)$ is the unit vector defining the orientation of the individual dimer. The last term in Eq.~\eqref{eq:trunc} defines the gyrotropic response of the cavity with parameter $B$ playing the role of the effective magnetic field applied to the cavity mirrors. The Hamiltonian is written in the dipole gauge, and the term, proportional to $g$ is the polarization operator of the dimers.
The Hamiltonian in Eq.~\eqref{eq:trunc} is normalized to the cavity energy $\omega$. We also note that the denominator $\sqrt{N}$ in the term corresponding to light-matter coupling originates from the approximation of constant dimer density. Indeed, coupling strength depends on the mode amplitude, which is inversely proportional to the square root of the mode volume. Therefore, when we keep the density of dimers constant we need to increase the mode volume proportionally to $N$, and hence coupling strength scales as $1/\sqrt{N}$. At the same time, one could consider the regime of the constant mode volume (at least for the finite $N$): in this case there will be no additional scaling factor $N^{-1/2}$.

As can be seen, the gyrotropy term in the Hamiltonian proportional to $B$ couples cavity modes with linear polarization. It is known, that in the presence of the gyrotropy the non-degenerate eigenmodes of the cavity are those with specific circular polarization. We first apply a unitary transformation (details are presented in Supplementary Material) in order to diagonalize the cavity part of the Hamiltonian. The diagonalized cavity Hamiltonian then reduces to $\omega_r \hat{r}^{\dagger}\hat{r}+\omega_l\hat{l}^{\dagger}\hat{l}$ where $r,l$ correspond to right- and left-circularly polarized modes, $\hat{r},\hat{l}$ are standard bosonic annihilation operators, and $\omega_{r,l}=\sqrt{B^2+1}\pm B$. The operator of net angular momentum of the cavity modes which is sometimes referred to as photon spin angular momentum density~\cite{bliokh2015transverse}is given by $\hat{L}= \hat{r}^{\dagger}\hat{r}-\hat{l}^{\dagger}\hat{l}$. The Hamiltonian is then written as
\begin{equation}\label{eq:expansion}
    \hat{H} = \hat{H}_{Mech}+\hat{H}_{Dicke}
\end{equation}
where $\hat{H}_{Mech}= (2J)^{-1}\sum_i \hat{p}_{\phi_i}^2 $ corresponds to the mechanic kinetic energy of the dimers and 
\begin{align} \label{eq:Hfull_rotated}
&\hat{H}_{Dicke} = \frac{\Delta}{2}\sum\limits_{i=1}^{N}\hat{\sigma}_{z_i}+\omega_r\hat{r}^{\dagger}\hat{r}+\omega_l\hat{l}^{\dagger}\hat{l}\\ &-\frac{ig/\sqrt{8N}}{\sqrt{\omega_r+\omega_l}}\sum\limits_{i=1}^{N}\hat{\sigma}_{x_i}[\omega_r\hat{r}e^{i\varphi_i}-\omega_l\hat{l}e^{-i\varphi_i}] +\mathrm{h.c.} \\  &+\frac{g^2}{8}\left[1+\frac{1}{N}\sum\limits_{i\neq j}^{N}\hat{\sigma}_{x_i}\hat{\sigma}_{x_j}\cos(\varphi_i-\varphi_j)\right],
        \end{align}
describes the coupled cavity modes and internal degrees of freedom of the dimers. 

\begin{figure}[!h]
\label{fig:potential}
\centering
\includegraphics[width=0.5\textwidth]{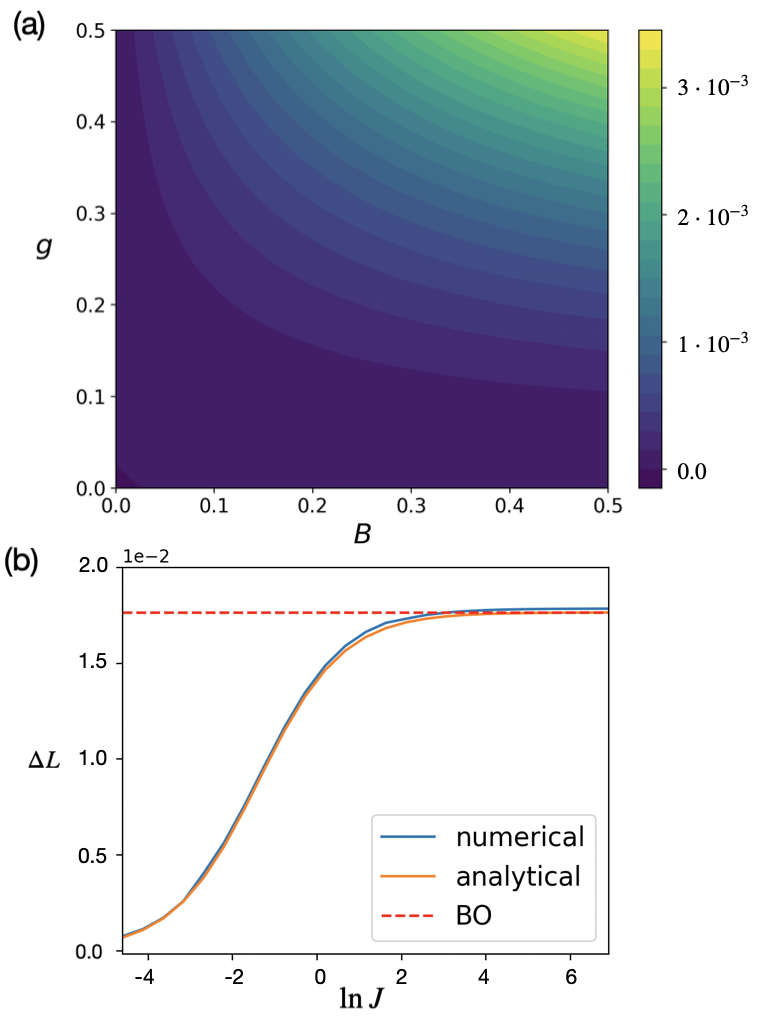}
\caption{(a) dependence of the optical angular momentum $L$ of a single dimer in the ground state on the magnetic field $B$ and coupling strength $g$. In the calculation $J=10^6$. (b) Dependence of the uncertainty of the angular momentum on the moment of inertia $J$. The dashed horizontal line corresponds to the result obtained within the Born-Oppenheimer approximation (BO). In the calculation $g=0.1,B=0.1$. The analytical result was computed with Eq.~\eqref{eq:DeltaL} }\label{fig:Fig2}
\end{figure}

We first note, that the total angular momentum of the system $\hat{\mathcal{L}}=\hat{L}+\sum_{i}p_{\varphi_i}$ is an integral of motion, $[\hat{\mathcal{L}},\hat{H}]=0$. Moreover, total angular momentum is a quantized quantity and the ground state of the system corresponds to $\mathcal{L}=0$, which is verified by the numerical solution of~\eqref{eq:Hfull_rotated}. In what follows we consider only $\mathcal{L}=0$ case. Thus, the total mechanical angular momentum is equal in amplitude and anti aligned with the optical angular momentum. In Fig.~\ref{fig:Fig2}(a) the dependence of the angular momentum on the coupling strength $g$ and magnetic field $B$ is plotted for the case of single dimer. 
It should be noted, that for any fixed value of $g$ the asymptotic of angular momentum at $B\rightarrow \infty$ is $1/B$ and for any fixed value of $B$ asymptotics for $g\rightarrow\infty$ is $e^{-g^2}$. The large $g$ and $B$ asymptotic analysis is presented in the Appendix B. 

Since total angular momentum $\mathcal{L}$ is the the integral of motion and it vanishes in the ground state, the uncertainties of the mechanical and optical angular momenta are the same. In Figure~\ref{fig:Fig2}(b) we plot the uncertainty of the optical angular momentum as a function of of the moment of inertia $J$ for the case of a single dimer. For small $J$ excitations of the mechanical degrees of freedom cost large energy ($1/J$) and thus the dimer localizes at the lowest rotational eigenstate and thus minimizes the uncertainty of the angular momentum. Since mechanical and and optical momenta a rigidly connected, the uncertainty of the optical angular momentum also vanishes. In the opposite limit of heavy dimers $J\rightarrow\infty$ the mechanical kinetic energy can be neglected and the uncertainty is defined by the uncertainty of the photon occupation numbers in the ground state of the Dicke Hamiltonian. For small $g$ one obtains the expressions for the angular momenta and its uncertainty (see the derivation  in the Appendix A):
\begin{align}
    L=\frac{g^2\Delta'}{8}\left(\frac{2\omega_r\omega_l+\Delta'(\omega_r+\omega_l)}{(\omega_r+\Delta')^2(\omega_l+\Delta')^2}\right)\frac{\omega_r-\omega_l}{\omega_r+\omega_l},
\end{align}
\begin{align}
   & \Delta L =\frac{g}{\sqrt{8(\omega_r+\omega_l)}}\sqrt{\frac{\omega^2_r}{(\Delta'+\omega_r)^2}+\frac{\omega^2_l}{(\Delta'+\omega_l)^2}}. \label{eq:DeltaL},
\end{align}
where $\Delta'=\Delta + \frac{1}{2J}$.
From Eq.~\eqref{eq:DeltaL} it can be seen that the uncertainty of the angular momentum decreases with the increase of total number of dimers $N$. Moreover, in the limit of large $J$, one can exploit the Born-Oppenheimer (BO) approximation, when one first omits the mechanical kinetic energy and finds the ground state of the Dicke Hamiltonian $E_0(\{\varphi\})$ with orientations of the dimers $\{\varphi\}$ treated as parameters. Afterwards one solves Schrodinger equation for the mechanical motion with $E_0(\{\varphi\})$. We see in Fig.~\ref{fig:Fig2}(b) that in large $J$ limit the BO approximation coincides with the numerical result. In what follows, we will resort to BO approximation since for the experimentally relevant situations, $J\gg 1$.

We now consider the effective inter-dimer interaction originating from the coupling to the cavity modes. We focus at the large $N$ limit describing the gas of dimers. It has been shown in~\cite{kudlis2023quantum} that one may use the $1/N$ expansion for the ground state energy of the Dicke Hamiltonian. Specifically, the leading correction to the ground state energy is given by the random phase approximation (RPA): 
\begin{align}
\delta E_0^{RPA} = \int_0^{\infty} \frac{\mathrm{d}\omega}{2\pi} \ln \left[ \mathrm{det} \left(1 - \frac{g^2}{4 N}D_{ph}(i\omega)\sum_{j=1}^{N}\Pi_{j}(i\omega)\right) \right], \label{Eq:RPA}
\end{align}
where $D_{ph}$ is the bare photon propagator and 

and $\Pi_{j}$ is the polarization bubble for the $j$~th dimer (expressions for $D_{ph}$ and $\Pi_j$ are presented in the Appendix).

The integral in Eq.~\eqref{eq:RPA} can be taken analytically, yielding,
\begin{align}
\delta E_0^{RPA}=\frac{1}{2}\sum_{l=1}^{4}\omega_{pol,l} -\frac{1}{2}(\omega_r+\omega_l) - \Delta,
\end{align}
where $\omega_{pol,l}$ are the energies of the polaritonic modes in the system which are found as the square roots $x_l$ of the zeros of the following fourth order polynomial presented in the Appendix 
In the absence of magnetic field, the roots can be written compactly yielding the following expression for the energy correction 
\begin{align}
&\delta E_0^{RPA}=\frac{\Delta+1}{2}\times\nonumber\\&\left[\sqrt{1+\frac{g^2\Delta(1+Z)}{4(\Delta+1)^2}}+\sqrt{1+\frac{g^2\Delta(1-Z)}{4(\Delta+1)^2}}-2\right], \label{eq:RPA}
\end{align}
where $Z=\left|\frac{1}{N}\sum_{j=1}^{N}e^{2i\varphi_j}\right|$. It can be seen from Eq.~\eqref{eq:RPA} that the dependence of the ground state energy on the dimer orientation appears only in the term proportional to $g^4$. The perturbation expansion in $g$ yields the following expression for the orientation dependent term:
\begin{align} \label{RPA:ans}
   & \delta E_0(\{\phi\})=\nonumber\\&-\frac{g^4\Delta^2}{64N^2}\frac{2\omega_r\omega_l+\Delta(\omega_r+\omega_l)}{(\omega_r+\omega_l)(\omega_r+\Delta)^2(\omega_l+\Delta)^2}\sum\limits_{i\neq j}\cos^2(\phi_i-\phi_j),
\end{align}
which for $B=0$ coincides with the leading term in the expansion over $g$ of Eq.~\eqref{eq:RPA}.

\begin{figure}[!h]
\label{fig:potential}
\centering
\includegraphics[width=0.5\textwidth]{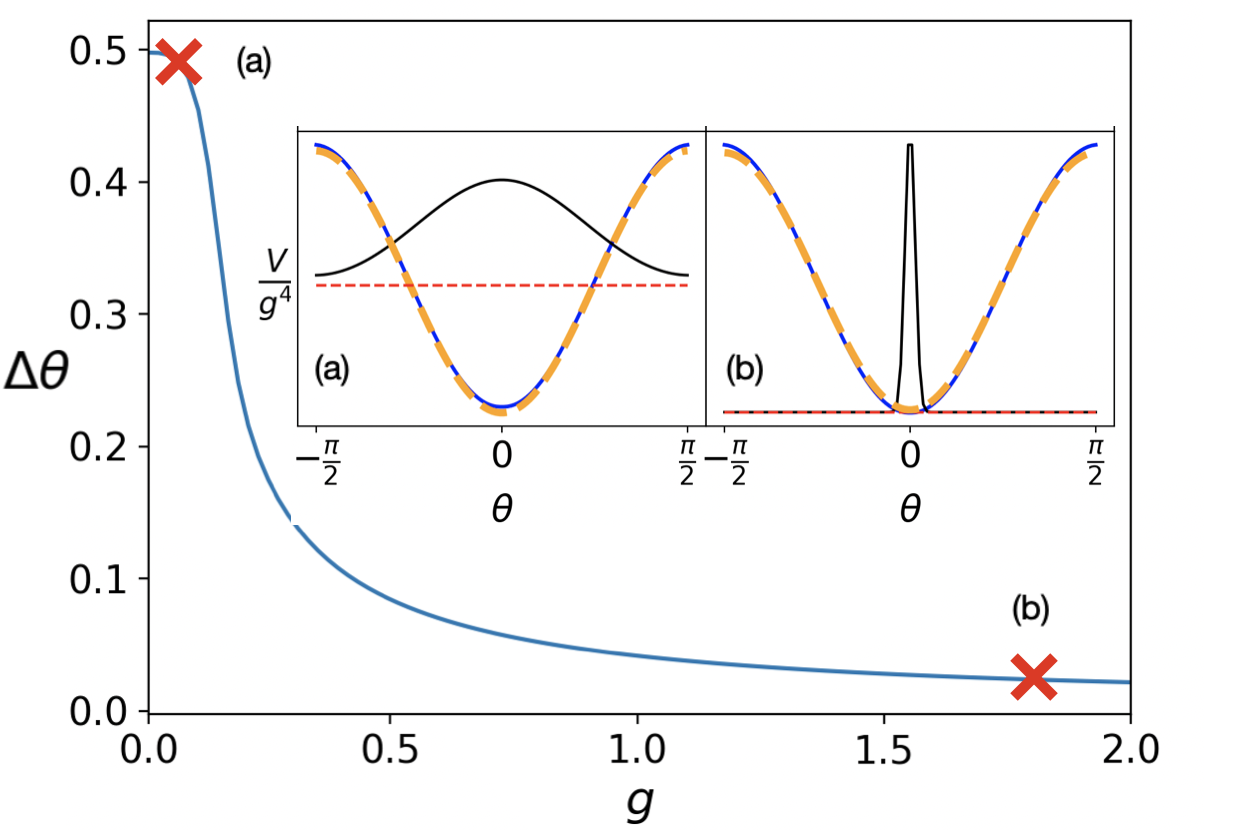}
\caption{The angle dispersion as the function of coupling strength $g$. Here it is assumed $\Delta=1, J=10^{7}, B=0.3$. On the subplot there is the profile of $V(\theta)$ potential (blue line), the profile of the energy correction in RPA (dashed orange line), energy level of the ground state (red dashed line) and the profile of the wave function of the ground state (black line).}\label{fig:potential}
\end{figure}
As can be seen, for $J\rightarrow \infty$ the ground state of the system corresponds to the fully aligned dimers, where as for finite $J$ one needs to compute the ground state of $\hat{H}_{Mech}+\delta E_0(\{\varphi\})$ to find the mechanical ground state of the system. For the case of $N=2$ this problem reduces to solving a one-dimensional Schrodinger equation for the wavefunction over relative orientation. The Schrodinger equation is just a Mathieu equation~\cite{arscott2014periodic} and its ground  solutions for different values of $g$ are shown in the insets of Fig.~\ref{fig:potential}. As $g$ is increased for the fixed $J$ the wavefunction becomes more localized and thus the uncertainty of the relative orientation decreases as shown in Fig.~\ref{fig:potential}.

We note that the orientation-dependent energy correction in the regime of finite density computed within RPA in Eq.~\eqref{eq:RPA} does not scale with $N$ and thus is not an extensive quantity and can not change the mechanical state of the dimers in the thermodynamic limit $N\rightarrow\infty$. Moreover, even if we consider the case of finite volume, i.e. make substitution $g\rightarrow g\sqrt{N}$ the energy correction still scales only as $\sqrt{N}$ and thus is sub-extensive. This is in accordance with the previously obtained results~\cite{kudlis2023quantum} stating that in the thermodynamic limit $N\rightarrow\infty$ electron photon interaction with a single cavity mode can not alter the ground state of the electronic system for arbitrary light-matter coupling strength. At the same time for sufficiently heavy dimers $J\rightarrow\infty$, the cavity-induced Casimir torque can still induce the alignment the dimers in the cavity.

It is worth to estimate the feasibility of the discussed effect.  We  can consider a mid-IR frequency range for the cavity, where the ultrastrong light-matter coupling has been realized recently~\cite{arul2022giant}. We choose $\hbar\omega=100~$meV and take $g=0.1,\Delta=1,B=0$. In this case, the aligning potential will have a relative height of $1~\mu$eV for two dimers. In the regime of finite mode volume, the height of the potential scales as $N$, and thus for $N=100$ the value of $0.1$~meV can be reached. The height of the aligning potential should be compared to the thermal fluctuations, and it can be seen that with the chosen parameters, the alignment can be observed for temperatures below 100~mK, which seems feasible for state of the art experiments.

To conclude we have shown that the collective light-matter coupling of an ensemble of dimers in a two-mode cavity leads to the emergence of the Casimir torque forcing to align the dimers along a single direction. Moreover, in the case of chiral cavity, when there exist an energy splitting between the two circularly polarized modes, the light-matter interaction induces a torque leading to the coherent rotation of the ensemble. In the future, it is worth to consider the torque emerging between two dissimilar group of dimers, where as has been recently shown, the strong correlations beyond the RPA approximation may arise~\cite{PhysRevA.108.L051701}. It is also worth noting, that the effective Hamiltonian for the mechanical motion is that of a fully connected quantum rotor model which has recently drawn considerable attention in the context of the quantum criticality~\cite{PhysRevB.96.104432}. Therefore, we believe that the proposed effect could find its applications in the quantum simulations of the correlated phases with cavity embedded cold atoms. 

We are grateful to A. Kudlis for useful discussions.

\newpage
\onecolumngrid
\newpage
\section{Supplementary}
\tableofcontents
\newpage

\section{S1. Diagonalization of the empty cavity Hamiltonian. Bogolubov transformation}
In the main part of the article it is noticed that in order to fully diagonalize cavity Hamiltonian we need to implement the Bogolubov transformation taking into account all orders of field $B$. So the exact transformation is 
\begin{equation}
    \begin{aligned}
        U_{B}=\begin{pmatrix}
        u_1 & u_1 & -u_2 & -u_2\\
        iu_1 & -iu_1 & iu_2 & -iu_2\\ 
        -u_2 & -u_2 & u_1 & u_1 \\ 
        -iu_2 & iu_2 & -iu_1 & iu_1
    \end{pmatrix},
    \end{aligned}
\end{equation}
where
\begin{equation}
    u_1=\frac{2+\omega_r+\omega_l}{4\sqrt{\omega_r+\omega_l}}, \ u_2=\frac{(\omega_r-\omega_l)^2}{4\sqrt{\omega_r+\omega_l}(2+\omega_r+\omega_l)}.
\end{equation}
Here we suppose $\omega_{r,l}=\sqrt{1+B^2}\pm B$. Also note that $\omega_{r,l}$ are values that normalized to the frequency of incoming photons, so these values are dimensionless. 

\section{S2. Perturbation theory for the case of weak light-matter coupling.}
In this section we consider weak light matter coupling regime, where one can use the perturbation expansion in coupling constant $g$. The weak interaction is described by the relationship $g\ll \Delta$. In this case we can consider parts that are proportional to the coupling strength $g$ as some perturbation potentials and exploit the results from the perturbation theory. Note, that this problem is solved without taking into account the Born-Oppenheimer approximation. The Hamiltonian reads
\begin{equation}
\begin{aligned}
    &\hat{H} = \hat{H_0} + g\hat{V}_1+g^2\hat{V}_2, \\  \hat{H_0} = \sum\limits_{i=1}^{N}\frac{\hat{p}^2_{\phi_i}}{2J} \ + &\frac{\Delta}{2}\left(\sum\limits_{i=1}^{N}\hat{\sigma}_{z_i}\right)+\omega_r\hat{r}^{\dagger}\hat{r}+\omega_l\hat{l}^{\dagger}\hat{l},
\end{aligned}
\end{equation}
where 
\begin{equation}
    \begin{aligned} 
        &\hat{V}_1 = -\frac{i}{\sqrt{8N(\omega_r + \omega_l)}}\sum\limits_{i=1}^{N}\hat{\sigma}_{x_i}\left(\omega_r \hat{r}e^{i\hat{\phi}_i}-\omega_l\hat{l}e^{-i\hat{\phi}_i}\right) + h.c.; \\ &
        \hat{V}_2 = \frac{1}{8}\left(1+\frac{1}{N}\sum\limits_{i\neq j}^{N}\cos(\hat{\phi}_i-\hat{\phi}_j)\hat{\sigma}_{x_i}\hat{\sigma}_{x_j}\right). 
    \end{aligned}
\end{equation}
The unperturbed part can be easily diagonalised as 
\begin{equation}
\begin{aligned}
    &E_{n,m,M, \textbf{k}} = \sum\limits_{i=1}^{N}\frac{{k}^2_i}{2J}+\Delta M+\omega_rn+\omega_lm, \\ & \Psi_{n,m,M,\textbf{k}} = \left\{\prod\limits_{i=1}^{N}e^{ik_i\phi_i}\right\}|n\rangle_r|m\rangle_l|M\rangle.
    \end{aligned}
\end{equation}
Here we denote $|n\rangle_{r(l)}$ as the Fock state of the $r(l)$-mode, $M$ is the total spin projection on the $z$-axis and $|M\rangle$ is eigenstate of the operator $\sum_{i}\hat{\sigma}_{z_i}$, $k_i\in\mathbb {Z}$ is the eigenvalue for operator $\hat{p}_i$ and $\textbf{k}=(k_1,..,k_N)^{T}$. The matrix elements of the perturbation operators read
\begin{equation}
\begin{aligned}
    \langle n',m',\textbf{k}'|\hat{V}_1|n,m,\textbf{k}\rangle&=\frac{i}{\sqrt{8N(\omega_r+\omega_l)}}\left(\omega_l\sqrt{m}\delta_{n',n}\delta_{m',m-1}-\omega_r\sqrt{n+1}\delta_{n',n+1}\delta_{m',m}\right)\sum\limits_{i=1}^{N}\hat{\sigma}_{x_i}e^{-i\phi_i}\delta_{k'_i,k_i+1} \\ &+\frac{i}{\sqrt{8N(\omega_r+\omega_l)}}\left(\omega_l\sqrt{m+1}\delta_{m',m+1}\delta_{n',n}-\omega_r\sqrt{n}\delta_{n',n-1}\delta_{m',m}\right)\sum\limits_{i=1}^{N}\hat{\sigma}_{x_i}e^{i\phi_i}\delta_{k'_i,k_i-1}; \\
    \langle n',m',\textbf{k}'|\hat{V}_2|n,m,\textbf{k}\rangle&=\frac{1}{8}\left(1+\frac{1}{N}\sum\limits_{i\neq j}^{N}\cos(\phi_i-\phi_j)\hat{\sigma}_{x_i}\hat{\sigma}_{x_j}\right)\delta_{n,n'}\delta_{m,m'}\delta_{\textbf{k}',\textbf{k}},
    \end{aligned}
\end{equation}
where $\delta_{m',m+1}$ is Kronecker symbol. 
For the ground state of perturbed system we obtain the expansion
\begin{equation}
\begin{aligned}
    &E_0=E^{0}_0+g^2E^{2}_2+g^3E^{3}_0+g^4E^{4}_0+..., \ \quad E^{0}_0 = -\frac{\Delta}{2}N, \\
    &\Psi_{0}=\Psi^{(0)}_0+g\Psi^{(1)}_0+g^2\Psi^{(2)}_0 + g^3\Psi^{(3)}_0 + ..., \ \quad \Psi^{(0)}_0=\left\{\prod\limits_{i=1}^{N}e^{ik_i\phi_i}\delta_{k_i}\right\}|0\rangle_r|0\rangle_l\left|-\frac{N}{2}\right\rangle.
    \end{aligned}
\end{equation}
For the second correction we have 
\begin{equation}
    E^{2}_0 = \langle 0|\hat{V}_2|0 \rangle+\sum\limits_{s_1\neq 0}\frac{\left|\langle s_1|\hat{V}_1|0 \rangle\right|^2}{E^{0}_0 - E^{0}_{s_1}} =\frac{\Delta'}{8}\frac{2\omega_r\omega_l+\Delta'(\omega_l+\omega_l)}{(\omega_r+\omega_l)(\omega_r+\Delta')(\omega_l+\Delta')},   
\end{equation}
where a state $s_1=\{n,m,M,\textbf{k}\}$ is parameterised by the eigenvalues set and $\Delta' = \Delta + \frac{1}{2J}$. For the next correction we obtain
\begin{equation}
    E^3_{0}=\sum\limits_{s_1\neq 0}\frac{\langle 0|\hat{V}_2|s_1 \rangle\langle s_1|\hat{V}_1|0 \rangle+\langle 0|\hat{V}_1|s_1 \rangle\langle s_1|\hat{V}_2|0 \rangle}{E^{0}_0 - E^{0}_{s_1}}+\sum\limits_{s_1,s_2\neq 0}\frac{\langle 0|\hat{V}_1|s_1 \rangle\langle s_1|\hat{V}_1|s_2 \rangle\langle s_2|\hat{V}_1|0\rangle}{(E^{0}_0 - E^{0}_{s_1})(E^{0}_0 - E^{0}_{s_2})}=0.
\end{equation}
The most interesting part appears in the forth part, i.e the dependence on phases.
\begin{equation}
\begin{aligned}
E^4_{0}&=\sum\limits_{s_1,s_2,s_3\neq 0}\frac{\langle 0|\hat{V}_1|s_1 \rangle\langle s_1|\hat{V}_1|s_2 \rangle\langle s_2|\hat{V}_1|s_3\rangle\langle s_3|\hat{V}_1|0\rangle}{(E^{0}_n - E^{0}_{s_1})(E^{0}_n - E^{0}_{s_2})(E^{0}_n - E^{0}_{s_3})}-\langle 0|\hat{V}_2|0 \rangle\sum\limits_{s_1\neq 0}\frac{\left|\langle s_1|\hat{V}_1|0 \rangle\right|^2}{(E^{0}_0 - E^{0}_{s_1})^2}-\sum\limits_{s_1\neq 0}\frac{\left|\langle s_1|\hat{V}_1|0 \rangle\right|^2}{(E^{0}_0 - E^{0}_{s_1})^2}\sum\limits_{s_1\neq 0}\frac{\left|\langle s_1|\hat{V}_1|0 \rangle\right|^2}{E^{0}_0 - E^{0}_{s_1}} \\ &+\sum\limits_{s_1\neq 0}\frac{\langle 0|\hat{V}_2|s_1 \rangle\langle s_1|\hat{V}_1|s_2 \rangle\langle s_2|\hat{V}_1|0 \rangle+\langle 0|\hat{V}_1|s_1\rangle\langle s_1|\hat{V}_2|s_2 \rangle\langle s_2|\hat{V}_1|0 \rangle+\langle 0|\hat{V}_1|s_1\rangle\langle s_1|\hat{V}_1|s_2 \rangle\langle s_2|\hat{V}_2|0 \rangle}{(E^{0}_n - E^{0}_{s_1})(E^{0}_n - E^{0}_{s_2})}+\sum\limits_{s_1\neq 0}\frac{\left|\langle s_1|\hat{V}_2|0 \rangle\right|^2}{E^{0}_0 - E^{0}_{s_1}}.
\end{aligned}
\end{equation}
From here we obtain a part without phases - $A$ and with phases:
\begin{equation}
    E^{4}_0=A-\frac{\Delta'^2}{64N^2}\frac{2\omega_r\omega_l+\Delta'(\omega_r+\omega_l)}{(\omega_r+\omega_l)(\omega_r+\Delta')^2(\omega_l+\Delta')^2}\sum\limits_{i\neq j}\cos^2(\phi_i-\phi_j)
\end{equation}
As we can the part with phases is non-positive, it means that the minimum is when all phases are equal, i.e all qubits tends to rotate in-phase.

Speaking about angular momentum of the system we can find it in the second order in expansion on $g$, i.e.
\begin{equation}
    L\approx\langle \Psi^{(0)}_0+g\Psi^{(1)}_0|\hat{L}|\Psi^{(0)}_0+g\Psi^{(1)}_0 \rangle=\frac{g^2\Delta'}{8}\left(\frac{2\omega_r\omega_l+\Delta'(\omega_r+\omega_l)}{(\omega_r+\Delta')^2(\omega_l+\Delta')^2}\right)\frac{\omega_r-\omega_l}{\omega_r+\omega_l}
\end{equation}
and dispersion 
\begin{equation}
    \Delta L \approx \frac{g}{\sqrt{8(\omega_r+\omega_l)}}\sqrt{\frac{\omega^2_r}{(\Delta'+\omega_r)^2}+\frac{\omega^2_l}{(\Delta'+\omega_l)^2}}.
\end{equation}
As it can be seen it is proportional to the difference between two energies $\omega_r-\omega_l=2B$, i.e the existence of the angular momentum is the result of the presence nonzero $B$ in the system.  On the other hand, if there is tremendous field $B$ in the system, then the angular momentum decreases as $L\approx \frac{g^2}{16B}$. As the result the maximum angular momentum is reached somewhere in the middle. As it can be seen from Fig.~\ref{fig:small_g} even for not very small coupling strength $g=0.5$ the angular momentum is still very low, i.e. less than $0.005$. 
\begin{figure}[t]
\label{fig:small_g}
\centering
\includegraphics[width=0.4\textwidth]{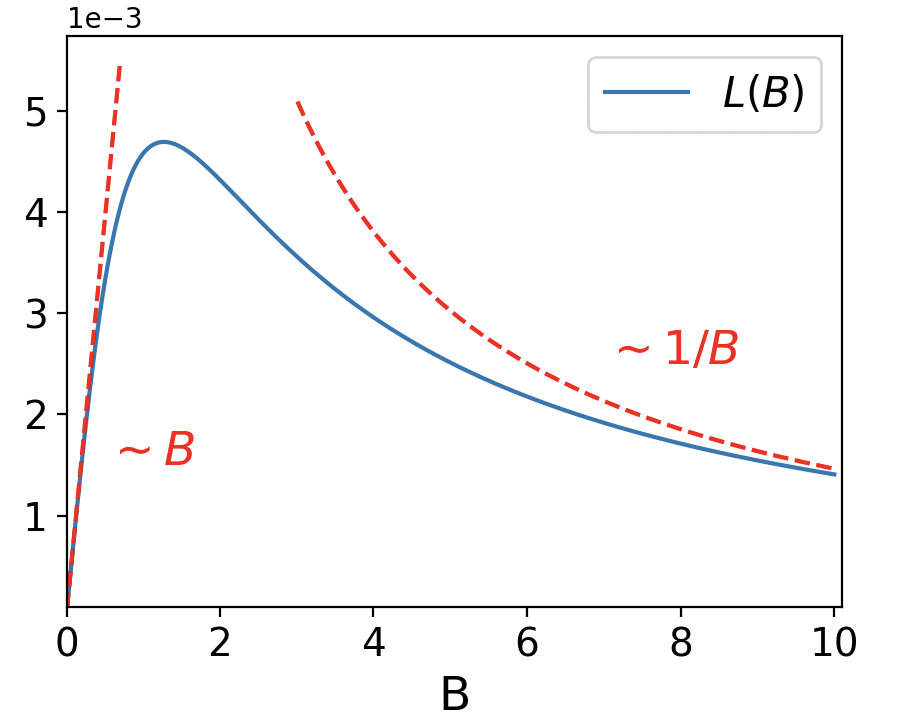}
\caption{The angular momentum as a function of the field $B$. We supposed here the incoming photon is resonant with the ground state to excited state transition of qubits, i.e $\Delta=1$ and fixed the coupling strength $g=0.5$.
}\label{fig:small_g}
\end{figure}

\section{S3. Perturbation theory for the case of strong light-matter coupling}
In this section we  consider the strong interaction $\Delta<g$ in the presence of the weak field $B\ll g$. Such regime can be realized in practice for relatively small detuning $\Delta$.  In this situation it is possible to  consider part that is proportional to the total $z$-spin projection as a perturbation. For simplicity we demonstrate the derivation for $N=1$ qubit in details. The Hamiltonian after transformation will be 
\begin{equation}
    \hat{H}_{N=1}=\frac{\left(\hat{p}_{\phi}-\hat{L}\right)^2}{2J}+\frac{\Delta}{2}\hat{\sigma}_z+\omega_r\hat{r}^{\dagger}\hat{r}+\omega_l\hat{l}^{\dagger}\hat{l}-\frac{ig}{\sqrt{8(\omega_r+\omega_l)}}\hat{\sigma}_x\left(\omega_r(\hat{r}-\hat{r}^{\dagger})-\omega_l(\hat{l}-\hat{l}^{\dagger})\right)+\frac{g^2}{8}.
\end{equation}
Then we implement the Born-Oppenheimer approximation, so we omit the part that  corresponds to the mechanical motion and also we make a rotation such that $\hat{\sigma}_x\rightarrow\hat{\sigma}_z$, $\hat{\sigma}_z\rightarrow -\hat{\sigma}_x$. After that we obtain
\begin{equation}
    \hat{H}^{N=1}_{Dicke}=\hat{H}_0+\hat{V}=\omega_r\hat{r}^{\dagger}\hat{r}+\omega_l\hat{l}^{\dagger}\hat{l}-\frac{ig}{\sqrt{8(\omega_r+\omega_l)}}\hat{\sigma}_z\left(\omega_r(\hat{r}-\hat{r}^{\dagger})-\omega_l(\hat{l}-\hat{l}^{\dagger})\right)+\frac{g^2}{8} -\frac{\Delta}{2}\hat{\sigma}_x,
\end{equation}
where $\hat{V}=-\frac{\Delta}{2}\hat{\sigma}_z$. The unperturbed part can be easily diagonilized as
\begin{equation}
\begin{aligned}
    E_{n,m}=\omega_rn+\omega_lm, \ & \Psi_{n,m}=\left\{|\uparrow\rangle|n\rangle_{-iq}|m\rangle_{iq}, \ |\downarrow\rangle|n\rangle_{iq}|m\rangle_{-iq}\right\}; \\ 
    & q=\frac{g}{\sqrt{8(\omega_r+\omega_l)}},
\end{aligned}
\end{equation}
where $|n\rangle_r$ denotes the Fock state of the displaced oscillator on $r$. As we can see in unperturbed system every level is twice degenerate. The existence of the perturbation removes the degeneracy and then we obtain
\begin{equation}
    \begin{aligned}
        E_{n,m,\pm}=\omega_rn+\omega_lm\mp\frac{\Delta}{2}e^{-4q^2}L_n(4q^2)L_m(4q^2), \ & \Psi^{(0)}_{n,m,\pm}=\frac{1}{\sqrt{2}}(|\uparrow\rangle|n\rangle_{-iq}|m\rangle_{iq} \pm |\downarrow\rangle|n\rangle_{iq}|m\rangle_{-iq}),
    \end{aligned}
\end{equation}
where $L_n(x)$ is the Lauguere polynom. So the ground energy is 
\begin{equation}
    E^{0}_0=-\frac{\Delta}{2}e^{-4q^2}, \ \Psi^{0}_0=\frac{1}{\sqrt{2}}(|\uparrow\rangle|0\rangle_{-iq}|0\rangle_{iq} + |\downarrow\rangle|0\rangle_{iq}|0\rangle_{-iq}).
\end{equation}

Just before to go further it is wise to make a comment. The use in such way of the perturbation theory is only valid when splitting energy levels by perturbation is small comparing to main levels, i.e $\omega_l \gg \Delta e^{-4q^2}$. It can be seduced if $q\gg 1$ that means $g \gg \sqrt{\omega_r+\omega_l}$ or simply $g\gg B$.

For our purposes it is sufficient to consider only the first order on $\Delta e^{-4q^2}$ in wavefunction to find nonzero value for angular momentum. So we introduce the expansion for the ground state
\begin{equation}
\begin{aligned}
&\Psi_0=\Psi^{(0)}_{0}+\Delta e^{-4q^2}\Psi^{(1)}_{0} +.., \\
&\Psi^{(1)}_{0} = \frac{1}{4}\sum\limits_{n\neq m}\frac{(2q)^{n+m}}{\sqrt{n!m!}(\omega_rn+\omega_lm)}\left[(1+(-1)^{n+m})\Psi^{(0)}_{n,m,+}-(1-(-1)^{n+m})\Psi^{(0)}_{n,m,-}\right].
\end{aligned}
\end{equation}
From here we can find the angular momentum up to the first order of $\Delta e^{-4q^2}$
\begin{equation}
L_1=\langle\Psi^{(0)}_0+g\Psi^{(1)}_0|\hat{L}|\Psi^{(0)}_0+g\Psi^{(1)}_0 \rangle\approx 4B\Delta q^2e^{-4q^2}.
\end{equation}
\begin{equation}
\Delta L_1\approx \sqrt{2(\omega_r+\omega_l)\Delta}qe^{-2q^2}.
\end{equation}
Again as we see the occurrence $L$ is a consequence of existence of the field $B$ in the system. Also in the case of tremendous value of the coupling strength $g$ and some fixed field $B$ the angular momentum will decrease rapidly. This fact is also a hint that the maximal possible value is somewhere in the middle Fig.~\ref{fig:big_g}. 

\begin{figure}[t]
\label{fig:big_g}
\centering
\includegraphics[width=0.8\textwidth]{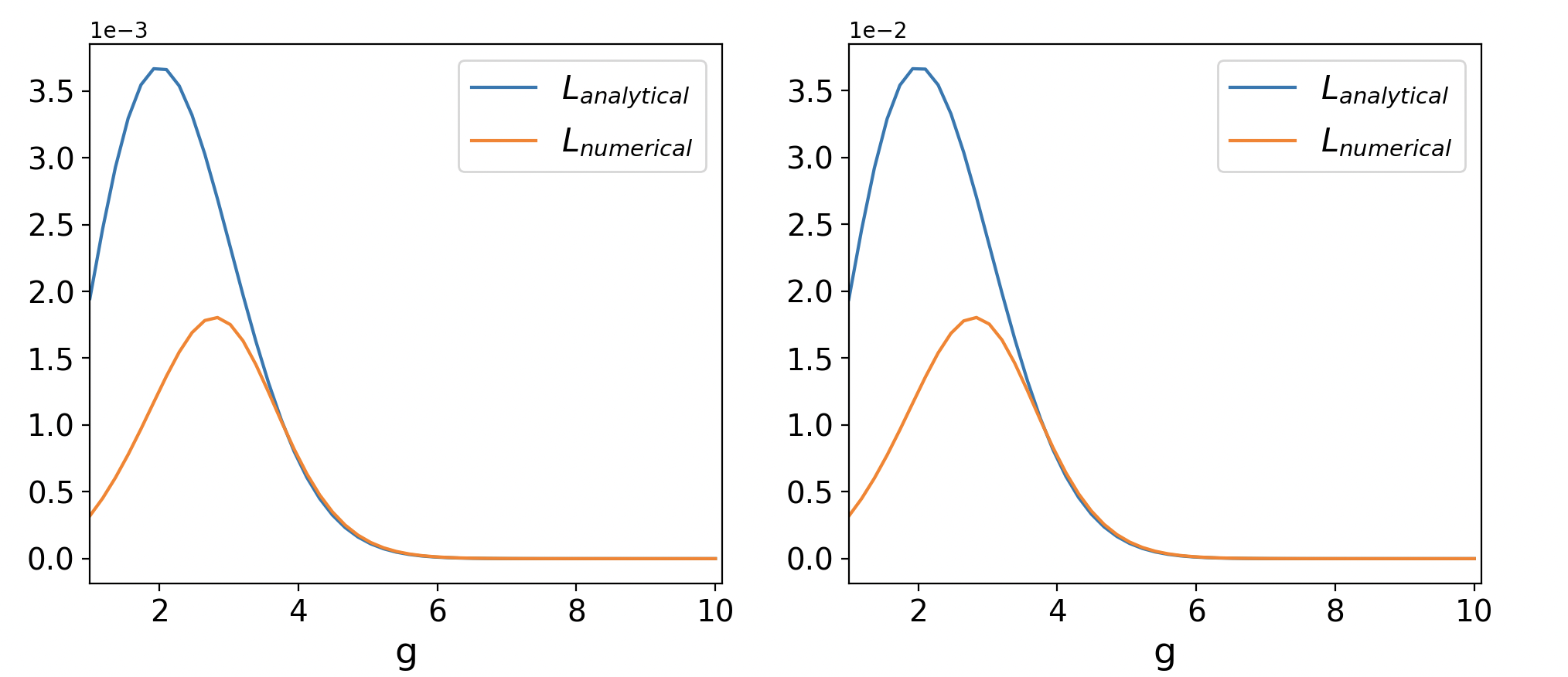}
\caption{Charts show the comparing between analytically calculated angular momentum for the $N=1$ qubit as a function of coupling strength $g$ and numerically for the case $B=0.01$ (left chat) and $B=0.1$ (right chat), for borh $\Delta=1$. As it can be seen the angular momentum is proportional to $B$ even for big $g$. The biggest difference between analytical way and numerical occurs in the vicinity of the maximal angular momentum.
}\label{fig:big_g}
\end{figure}

Such derivation can be realized for any number of qubits in the system. Here we also represent the analytically counted angular momentum for the system with $N=2$ qubits.
\begin{equation}
    L_2\approx 4B\Delta q^2e^{-2q^2}
\end{equation}
Furthermore it is proved numerically that in systems with $N$ qubits we will have $L_N\approx 4B\Delta q^2e^{-\frac{4}{N}q^2}$.

We now consider an intermediate regime, when  the splitting of energy levels due to the spin $\hat{V}=-\Delta\hat{S}_x$ and energy levels of $l-$mode are very close to each other such that some energy levels become degenerated. For example, in the system with $N=1$ qubit it means $\omega_l\approx \Delta e^{-4q^2}$ Fig.~\ref{fig:levels}. This regime we call intermediate. Mathematically it means that we cannot use the perturbation theory as we did. Instead we need to honestly diagonalize matrix in the vicinity of level degeneracy.

\begin{figure}[t]
\label{fig:levels}
\centering
\includegraphics[width=0.8\textwidth]{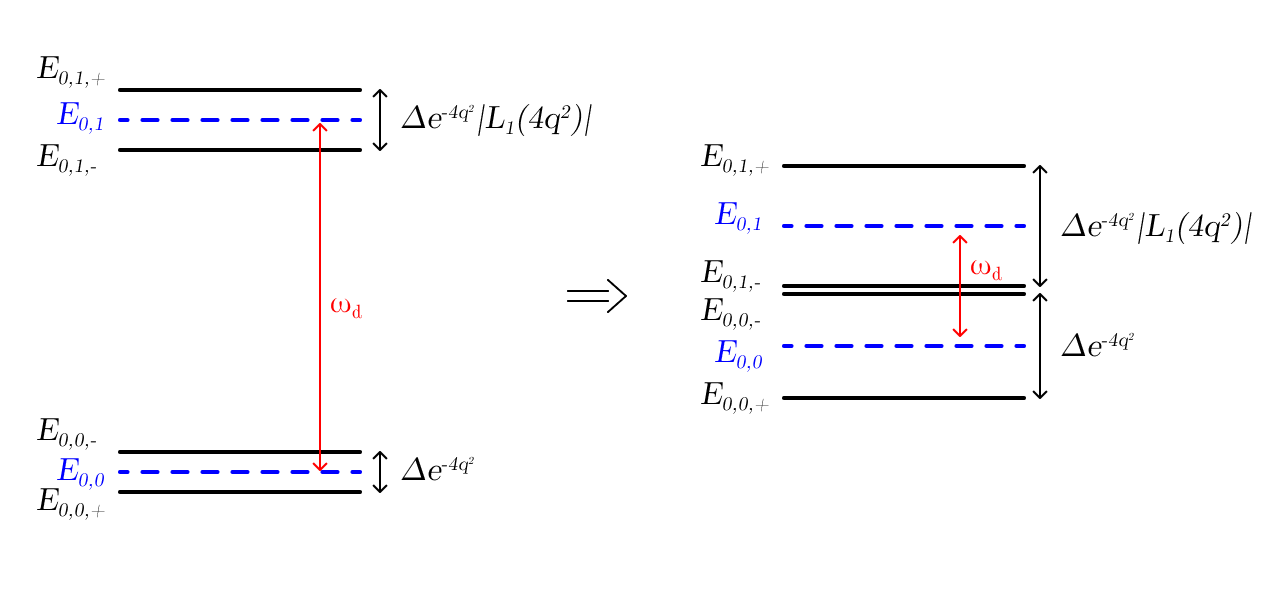}
\caption{The left part (II) region, i.e the splitting levels due to the $\Delta$ part is small comparing to the excitations of the $l$-mode. The right part is the resonance regime, i.e they are the same order.}\label{fig:levels}
\end{figure}

Now we are going to provide a full derivation for the system with $N=1$ qubit taking into account an explicit diagonalisation only the zeroest and the first excitation of $d$-mode. So in order to find the ground state we should find the minimal eigenvalue of the matrix
\begin{equation}
    \begin{aligned}
         H = \begin{pmatrix}
        0 & -\frac{\Delta}{2}e^{-4q^2} & 0 & i\Delta q e^{-4q^2}\\
        -\frac{\Delta}{2}e^{-4q^2} & 0 & -i\Delta q e^{-4q^2} & 0\\ 
        0 & -i\Delta q e^{-4q^2} & \omega_l & -\frac{\Delta}{2}e^{-4q^2}L_1(4q^2) \\ 
        i\Delta q e^{-4q^2} & 0 & -\frac{\Delta}{2}e^{-4q^2}L_1(4q^2) & \omega_l
    \end{pmatrix}
    \end{aligned}
\end{equation}
Previously we omitted the excited levels of the $l$ mode. Now we consider the case when the energy of the first excited state of the $l$ mode is small enough such that
\begin{equation}
    \omega_l+\frac{\Delta}{2}e^{-4q^2}L_1(4q^2)\approx\frac{\Delta}{2}e^{-4q^2} \ \leftrightarrow \ \omega_l=2q^2\Delta e^{-4q^2}.
\end{equation}
This relation is in a good match with the results from the straightforward numerical modulation of the system Fig.~\ref{fig:double}(a). 
After that we can find the ground energy as $E_0=-\frac{\Delta}{2}e^{-4q^2}\sqrt{1+4q^2}$ and the state as 
\begin{equation}
\begin{aligned}
&\Psi_0=\frac{q}{\sqrt{1+4q^2+\sqrt{1+4q^2}}}\left\{-|\uparrow\rangle|0\rangle_{-iq}|1\rangle_{iq}+|\downarrow\rangle|0\rangle_{iq}|1\rangle_{-iq}+\frac{1+\sqrt{1+4q^2}}{2q}|\uparrow\rangle|0\rangle_{-iq}|0\rangle_{iq}+\frac{1+\sqrt{1+4q^2}}{2q}|\downarrow\rangle|0\rangle_{iq}|0\rangle_{-iq}\right\}.
    \end{aligned}
\end{equation}
Using this state we can the angular momentum
\begin{equation}\label{eq:L1}
L_1\approx\langle\Psi_0|\hat{L}|\Psi_0\rangle=\frac{2q^2}{1+\sqrt{1+4q^2}}.
\end{equation}

\begin{figure}[t]
\label{fig:double}
\centering
\includegraphics[width=0.9\textwidth]{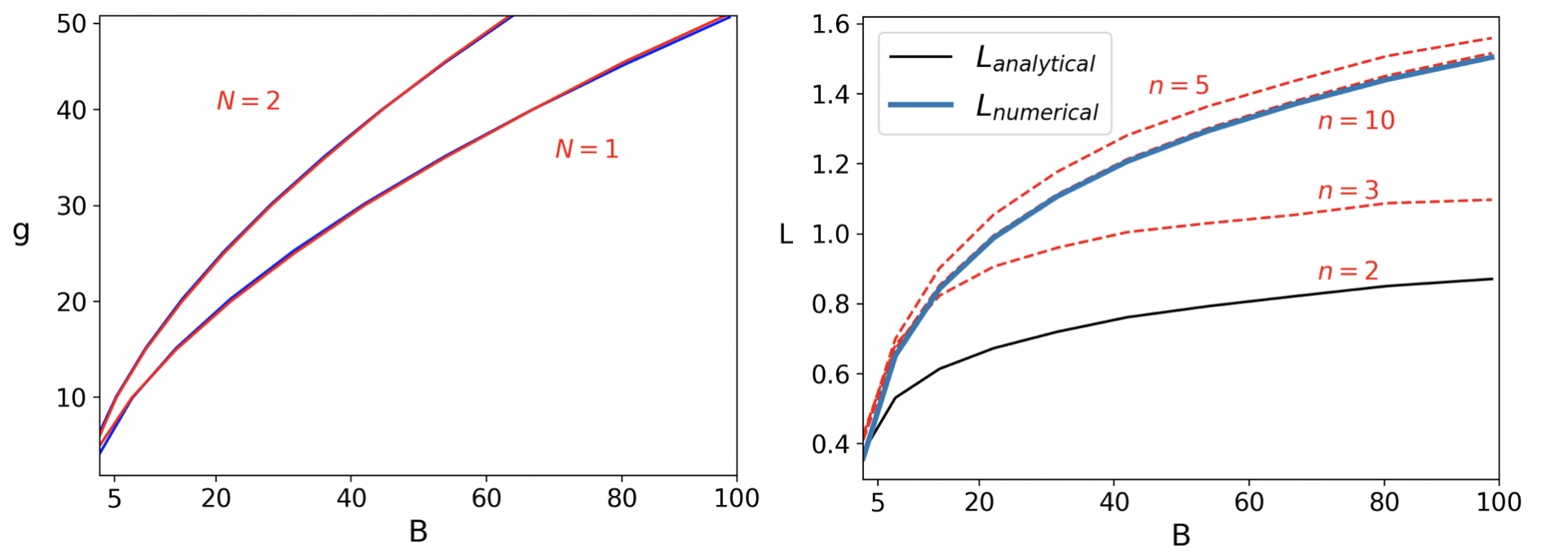}
\caption{(a) The dependence between field $B$ and coupling strength $g$ assumimg the incoming photons are in resonant with qubits, i.e. $\Delta=1$. Red lines correspond to the analytical resonance relationships for $N=1$ and $N=2$ respectively, blue lines correspond to the numerical solution. (b) The maximal possible angular momentum for the resonant photons, i.e. $\Delta=1$ as a dependence on field $B$ for $N=1$ system. The blue line correspond to numerical solution, the black line is the simplest analytical model ~\eqref{eq:L1} and red lines are solution of ~\eqref{eq:LN} taking into account interaction between first $n-1$ excitations.}\label{fig:double}
\end{figure}

As it can be seen this estimation is already gives huge values such as $L_1>0.1$. Please note, that this formula can't be used for very small fields $B$, because the angular momentum does not vanish. Nevertheless, the numerical model predicts that the real values of $L$ are larger than this estimation and so it is not sufficient to achieve the precise result taking into account only the interaction of the zeroest and first excitation. However, this procedure can be spread to the larger number of excitation. Indeed, let the ground state is described by mixing first $n-1$ excitations and the ground state:
\begin{equation}
    \Psi_0\approx\sum\limits_{j=0}^{n-1}C^{\uparrow}_j|\uparrow\rangle|0\rangle_{-iq}|j\rangle_{iq}+C^{\downarrow}_j|\downarrow\rangle|0\rangle_{iq}|j\rangle_{-iq},
\end{equation}
where $C^{\uparrow;\downarrow}_i$ are complex normalised coefficients, so $|\Psi_0|^2=1$. After some algebra we can obtain 
\begin{equation}\label{eq:LN}
L_1\approx\langle\Psi_0|\hat{L}|\Psi_0\rangle=-\sum\limits_{j=1}^{n-1}\left\{j\left[|C^{\uparrow}_j|^2+|C^{\downarrow}_j|^2\right]+2q\sqrt{j}\left[C^{\uparrow *}_jC^{\uparrow}_{j+1}-C^{\downarrow *}_jC^{\downarrow}_{j-1}\right]\right\}.
\end{equation}
Then using this formula we can calculate corrections to angular momentum Fig.~\ref{fig:double}(b). As it can be seen it sufficient to predict the real value of the angular momentum considering mixing of the first 9 excitations and the ground state. 

Despite the fact that it is possible to find an intermediate solution for any large $(g,B)$ for the fixed $\Delta$, there are several restrictions on the occurence of intermediate regime in the systems with small $g,B$. Indeed, the two main relationships should be preserved in order to observe intermediate solution, i.e the condition of appliance of the perturbation theory and degeneracy of level splitting due to the spin perturbation potential. Altogether it leads to the more interpreted condition, i.e. $q\geq 1/2$. So it can be rewritten as
\begin{equation}
\begin{aligned}
    &\frac{g^4}{16}\geq 1+B^2, \\
    &\Delta\geq 2e\omega_l.
\end{aligned}
\end{equation}
The first inequality means that the intermediate solution takes place for any field $B$ but only for coupling strength that are greater than 2, i.e $g\geq 2$. This is the main problem to find such state in real systems because of practical difficulties to achieve such strength in real systems. The second inequality means that not for all small frequencies it is possible to observe such regime. This happens mostly because of restriction on $g$.

For the system with $N=2$ qubits it is only possible to write analytically the intermediate condition Fig.~\ref{fig:double}(a)
\begin{equation}
    \omega_l\approx\Delta e^{-2q^2}.
\end{equation}
Nevertheless we can say that for the system with $N$ qubits when all of them are in-phase, i.e $\phi_i=\phi_j$ for any $i,j$ and fixed field $B$, angular moments are connected by relationship $L_N=\sqrt{N}L_1$. This relation proven numerically for several systems. Also it implies an interesting thing. We can achieve the target angular momentum by decreasing field $B$ and increasing coupling strength $g$ and number of qubits $N$ in the system. This effect might be very useful in future applications. 

The intermediate state is also remarkably by the fact that the angular momentum attains its maximal value in this state with fixed two parameters, i.e $\Delta, B$ or $\Delta, g$. In Fig. ~\ref{fig:Lnum} we calculated numerically the angular momentum for different $B, g$ with fixed $\Delta=1$ for $N=1$ (a) and $N=2$ (b). As it can be seen the maximal values on verticals or horizontals are achieved in intermediate state.

\begin{figure}[t]
\label{fig:Lnum}
\centering
\includegraphics[width=0.9\textwidth]{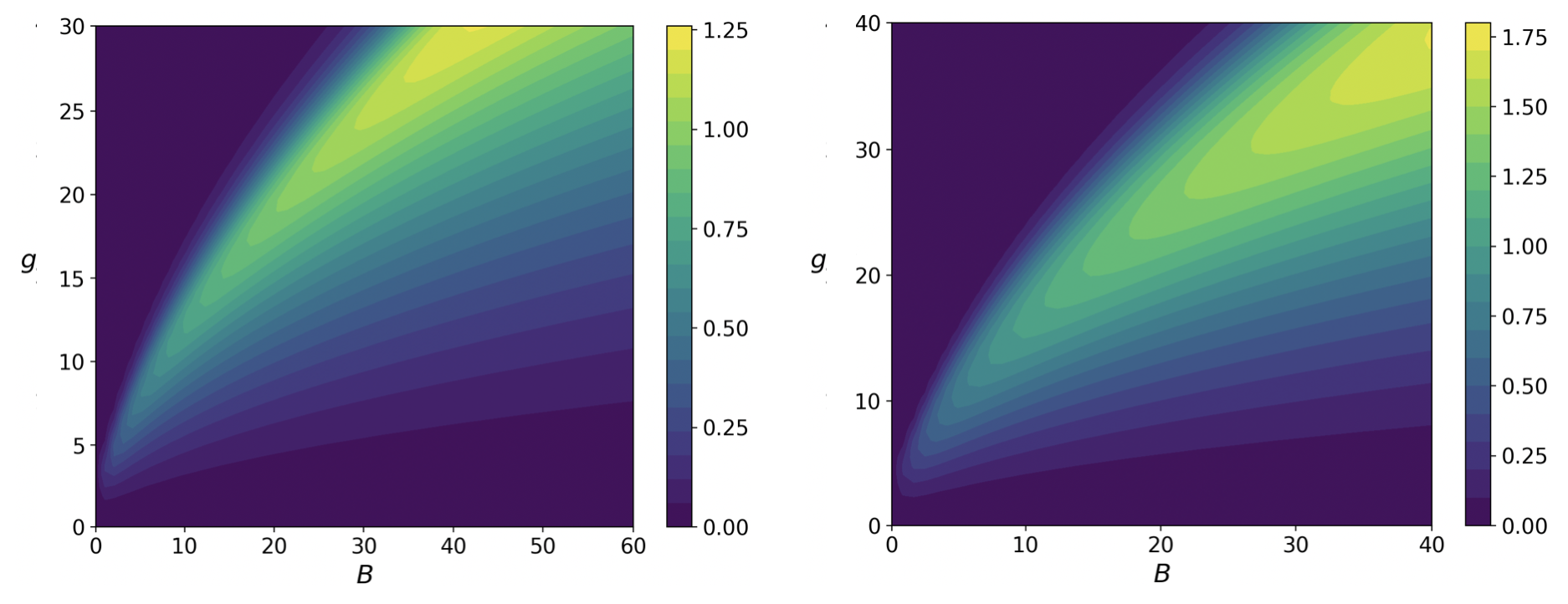}
\caption{The numerical calculated angular momentum for $N=1$ (a) and $N=2$ (b) qubits. It is assumed $\Delta=1$.}\label{fig:Lnum}
\end{figure}

\section{S4. Ground state energy correction within the Random Phase approximation}
In this section we derive the ground state energy correction due to the light-matter coupling in the limit of large number of dimers, $N\rightarrow\infty$. As has been shown in~\cite{kudlis2023quantum} this energy correction is governed by the random phase approximation. Specifically, 
\begin{align}
\delta E_0^{RPA} = \int_0^{\infty} \frac{\mathrm{d}\omega}{2\pi} \ln \left[ \mathrm{det} \left(1 - \frac{g^2}{4N}D_{ph}(i\omega)\sum_{j=1}^{N}\Pi_{j}(i\omega)\right) \right], \label{EqSup:RPA}
\end{align}
Where $D_{ph}$ is the matrix of bare photon propagator, and the $\Pi_j$ is the polarizability matrix of $j$-th qubit.
The photon propagator can be obtained from the  equations of motion for the photon cananonical coordinate and momentum~\cite{sedov2022cavity}. In the basis of circularly polarized modes, the propagator takes the diagonal form:
\begin{align}
D_{ph}(\omega) = -\begin{pmatrix} \frac{\omega^2}{(\omega-\omega_r)(\omega+\omega_l)} & 0 \\ 0 & \frac{\omega^2}{(\omega+\omega_r)(\omega-\omega_l)} \end{pmatrix}.
\end{align}
The qubit polarizability is most readily written in the cartesian basis. Polarizability $\Pi_j$ of the $j$-th qubit is a bare bubblue with two vertices corresponding to coupling to $x$ and $y$ polarized modes. The strength of coupling is proportional to the orientation of dimer $\mathbf{n}_j$. Thus, in the cartesian basis, the polarizability matrix can be written as:
\begin{align}
\Pi_j(\omega) = -\frac{2\Delta}{\omega^2-\Delta^2}\begin{pmatrix} \cos^2\varphi_j & \cos\varphi_j\sin\varphi_j \\ \cos\varphi_j\sin\varphi_j & \sin^2\varphi_j  \end{pmatrix}.
\end{align}
In the basis of circularly polarized modes, the polarizability matrix reads
\begin{align}
\Pi_j(\omega) = -\frac{\Delta}{\omega^2-\Delta^2}\begin{pmatrix} 1 & e^{-2i\varphi_j} \\ e^{2i\varphi_j} & 1  \end{pmatrix}.
\end{align}
The integral in Eq.~\eqref{EqSup:RPA} can be taken analytically yielding:
\begin{align}
\delta E_0^{RPA}=\frac{1}{2}\sum_{l=1}^{4}\omega_{pol,l} -\frac{1}{2}(\omega_r+\omega_l) - \Delta,
\end{align}
where $\omega_{pol,l}$ are the energies of the polaritonic modes in the system which are found as the square roots $x_l$ of the zeros of the following fourth order polynomial $P(x)$:
\begin{align}
P(x)=(x-\Delta^2)^2(x-\omega_r^2)(x-\omega_l^2)-\frac{g^2}{2}\Delta x(x-\Delta^2)(x-1)+\frac{g^4}{16}(1-|Z|^2)\Delta^2x^2,
\end{align}
where $Z=\frac{1}{N}\sum_{j=1}^{N}e^{i2\phi_j}$. The roots of the polynomial can be found in the absence of magnetic field, $\omega_r=\omega_l=1$ and the result is presented in Eq.~\eqref{RPA:ans} in the main text. 

\end{document}